\documentclass[a4paper,10pt]{article}

\def\dpa#1#2{\frac{\partial{#1}}{\partial#2}}

\def\equ#1{(\ref{#1})}
\def\be#1{\begin{equation}\label{#1}}
\def\ee{\end{equation}}
\def\vet#1{\mathbf{#1}}
\makeatletter
\@addtoreset{equation}{section}
\makeatother

\title{On the Laplace equation in d-dimension}
\author{R. R. Landim\thanks{email: renan@fisica.ufc.br}\\ Departamento de F\'{\i}sica, \\ Universidade Federal do Cear\'a, \\ Caixa Postal 6030,
60455-900, Fortaleza, Cear\'a, Brazil}

\begin{document}

\maketitle

\begin{abstract}
We develop a method to obtain the general solution of the Laplace equation in $d$-dimension in ultraspherical coordinates.
\end{abstract}

%\section{Spherical solutions}
The coordinates of a point in $d$-dimensional space are given by the position vector $\vet{r}_d=(x_d,x_{d-1},\dots,x_1)$, $|\vet r_d|=r$. We define the ultraspherical coordinates by $\vet{r}_j=r_j(cos\theta_j\hat{\vet k}_j+\sin\theta_j\hat{\vet r}_{j-1})=x_j\hat{\vet k}_j+\vet{r}_{j-1}$, $0\le\theta_j\le\pi$, $j=3,\cdots ,d$, $\vet{r}_2=r_2\cos\phi\hat{\vet k}_1+r_2\sin\phi\hat{\vet k}_2, 0\le\phi\le2\pi$, $\hat{\vet k}_i\cdot \hat{\vet k}_j=\delta_{ij}$. The hat symbol over a bold letter always means a unitary vector.

We want solutions that are finite in the angles and the frontier conditions are radial only and the domain in the angles are complete, i.e, $0\le\phi\le2\pi$ and $0\le\theta_j\le\pi$, $j=3,\cdots ,d$.

\section{The laplacian operator}
Before to find the solution of the Laplace equation in ultraspherical coordinates, we want to find the explicit form of the laplacian operator. Consider the operator
\be{oper1}
\dpa{^2}{z^2}+\frac{1}{\rho^\alpha}\dpa{}{\rho}\left(\rho^\alpha\dpa{}{\rho}\right)=\dpa{^2}{z^2}+\dpa{^2}{\rho^2}+\frac{\alpha}{\rho}\dpa{}{\rho}.
\ee
Applying a coordinate transform $z=r\cos\theta$ and $\rho=r\sin\theta$:
$$
\dpa{}{\rho}=\sin\theta\dpa{}{r}+\frac{\cos\theta}{r}\dpa{}{\theta},
$$
$$
\dpa{^2}{z^2}+\dpa{^2}{\rho^2}=\frac{1}{r}\dpa{}{r}\left(r\dpa{}{r}\right)+\frac{1}{r^2}\dpa{^2}{\theta^2},
$$
then
$$
\dpa{^2}{z^2}+\dpa{^2}{\rho^2}+\frac{\alpha}{\rho}\dpa{}{\rho}=\dpa{^2}{r^2}+r\dpa{}{r}+\frac{1}{r^2}\dpa{^2}{\theta^2}+\frac{\alpha}{r\sin\theta}\left(\sin\theta\dpa{}{r}+\frac{\cos\theta}{r}\dpa{}{\theta}\right)\Rightarrow
$$
\be{relation1}
\dpa{^2}{z^2}+\frac{1}{\rho^\alpha}\dpa{}{\rho}\left(\rho^\alpha\dpa{}{\rho}\right)=\frac{1}{r^{\alpha+1}}\dpa{}{r}\left(r^{\alpha+1}\dpa{}{r}\right)+\frac{1}{r^2}\frac{1}{\sin^\alpha\!\theta}\dpa{}{\theta}\left(\sin^{\alpha}\!\theta\dpa{}{\theta}\right)
\ee
For $d=2$:
\be{lap2}
\nabla_2^2=\dpa{^2}{x^2}+\dpa{^2}{y^2}=\frac{1}{\rho}\dpa{}{\rho}\left(\rho\dpa{}{\rho}\right)+\frac{1}{\rho^2}\dpa{^2}{\phi^2}, \quad x=\rho\cos\phi,\quad y=\rho\sin\phi.
\ee
For $d=3$:
$$
\nabla_3^2=\dpa{^2}{z^2}+\nabla_2^2=\dpa{^2}{z^2}+\frac{1}{\rho}\dpa{}{\rho}\left(\rho\dpa{}{\rho}\right)+\frac{1}{\rho^2}\dpa{^2}{\phi^2}.
$$
Using \equ{relation1} with $\alpha=1$ and $z=r\cos\theta$ and $\rho=r\sin\theta$:
\be{lap3}
\nabla_3^2=\frac{1}{r^2}\dpa{}{r}\left(r^2\dpa{}{r}\right)+\frac{1}{r^2\sin\theta}\dpa{}{\theta}\left(\sin\theta\dpa{}{\theta}\right)+\frac{1}{r^2\sin^2\!\theta}\dpa{^2}{\phi^2}=\frac{1}{r^2}\dpa{}{r}\left(r^2\dpa{}{r}\right)+\frac{1}{r^2}\hat{L}_3.
\ee
By induction, supposing valid for $d-1$:
\be{lapdm1}
\nabla_{d-1}^2=\frac{1}{r_{d-1}^{d-2}}\dpa{}{r_{d-1}}\left(r_{d-1}^{d-2}\dpa{}{r_{d-1}}\right)+\frac{1}{r_{d-1}^2}\hat{L}_{d-1}.
\ee
Using \equ{relation1} with $\alpha=d-2$ and $x_d=r\cos\theta_d$ $r_{d-1}=r\sin\theta_d$ we have
$$
\nabla_d^2=\dpa{^2}{x_d^2}+\nabla_{d-1}^2=\dpa{^2}{x_d^2}+\frac{1}{r_{d-1}^{d-2}}\dpa{}{r_{d-1}}\left(r_{d-1}^{d-2}\dpa{}{r_{d-1}}\right)+\frac{1}{r_{d-1}^2}\hat{L}_{d-1}
$$
\be{lapd}
=\frac{1}{r^{d-1}}\dpa{}{r}\left(r^{d-1}\dpa{}{r}\right)+\frac{1}{r^2}\hat{L}_d
\ee
were
\be{ld}
\hat{L}_d=\frac{1}{\sin^{d-2}\!\theta_d}\dpa{}{\theta_d} \left(\sin^{d-2}\!\theta_d\dpa{}{\theta_d}\right)+\frac{1}{\sin^2\!\theta_d}\hat{L}_{d-1}
\ee
We see from \equ{lapd} that we can obtain the solution of the Laplace equation if we known the eigenfunctions of the angular operator $\hat{L}_d$. 
\section{The general solution}
Let the eigenfunctions of $\hat{L}_d$ given by $\Psi(\Omega)$, where $\Omega$ is the angles coordinates $\Omega=(\theta_d,\theta_{d-1},\cdots, \theta,\phi)$.
\be{auto-ld}
\hat{L}_d\Psi_\beta(\Omega)=\beta\Psi_\beta(\Omega).
\ee
Now supposing the solution of the Laplace equation by the separable form $\Phi_\beta(r,\Omega)=R_\beta(r)\Psi_\beta(\Omega)$, we get the radial equation:
\be{radiald}
\frac{1}{r^{d-1}}\frac{d}{dr}\left(r^{d-1}\frac{dR_\beta(r)}{dr}\right)+\frac{\beta}{r^2}R_\beta(r)=0.
\ee
This is a second order Euler equation whose solution is given by $r^\alpha$, where $\alpha$ is to be determined. We find
\be{alpha}
\beta=-\alpha(\alpha+d-2).
\ee
Then the general solution is the form
\be{generald}
\Phi(r,\Omega)=\sum_\alpha (A(\alpha)r^\alpha+B(\alpha)r^{-(\alpha+d-2)})\Psi_\alpha(\Omega),
\ee
where $\hat{L}_d\Psi_\alpha(\Omega)=-\alpha(\alpha+d-2)\Psi_\alpha(\Omega)$.

\section{The eigenfunctions of the angular operator}
Since we are interested in the general solutions that the frontier conditions are radial only, the angular part are the same for all type of frontier condition. Consider first the non-trival solution of \equ{generald} which is independent of the angles. Clearly a solution is for $\alpha=0$ with $A(0)=0$, that is $B(0)r^{-(d-2)}$. To construct a solution with an angular dependency, we observe that the laplacian operator is invariant under translation. Making a unitary translation in the $x_d$ direction, we conclude that
$$
\frac{1}{|\vet{r}_d-\vet k_d|^{d-2}}=\frac{1}{(1+r^2-2r\cos\theta_d)^{\frac{d-2}{2}}},
$$
is a solution of the Laplace equation for $r\ne1$ that goes to zero when $r\rightarrow\infty$.
We define the ultraspherical polynomials by
\be{ultra-esf}
\frac{1}{(1+r^2-2r\cos\theta_d)^{\frac{d-2}{2}}}=\sum_{l=0}^\infty r^l P_{l,d}(\cos\theta_d), \quad r<1.
\ee
Note that for $d=3$ the equation above gives the definition of the Legendre polynomials. 

Applying the laplacian operator \equ{lapd} in both sides of \equ{ultra-esf} and taking the fact that $r^l$ are linearly independent for each $l$ we have
\be{eq-ue}
\hat{L}_d P_{l,d}(\cos\theta_d)=-l(l+d-2)P_{l,d}(\cos\theta_d).
\ee
For $d=3$ the equation above reduces to the Legendre equation. Note that this eigenfunctions depend only of the angle $\theta_d$. This solutions can only be applied in problems that are independent of the other angles. 

To find a general solution for $d$-dimension, observe that a solution in $(d-1)$-dimension is also a solution in $d$-dimension. Consider first the case $d=3$ for a better understand. Is well known that $g_m=\rho^m e^{im\phi}$ is a solution for $d=2$. Then is also a solution for $d=3$. But linear combinations of this solutions can solve a part of  two dimensional problems only (finites for $\rho=0$). To find a tri-dimensional solution let us try
\be{fm}
f_m=\frac{g_m}{|\vet r-\vet k_3|^\alpha}, r\ne1,
\ee
where $\alpha$ is to be determined by the requirement that $f_m$ being a solution of the Laplace equation. Note that the dependency of tri-dimensional angle in the denominator and for $m=0$ and $\alpha=1$ we have a solution of the type \equ{ultra-esf} with $d=3$. Applying the laplacian operator in both side of \equ{fm} we have
$$
\nabla_3^2 f_m=2\vec{\nabla}_3 g_m\cdot\vec{\nabla}_3\frac{1}{|\vet r-\vet k_3|^\alpha}+g_m\nabla_3^2\frac{1}{|\vet r-\vet k_3|^\alpha}=0.
$$
since $\vec{\nabla}_3=\dpa{}{x_3}+\vec{\nabla}_2$ and $\vet r-\vet k_3=\vet r_2+(x_3-1)\vet k_3$, with $\rho=r_2=r\sin\theta$, $x_3=r\cos\theta$, we get 
$$
\vec{\nabla}_3\frac{1}{|\vet r-\vet k_3|^\alpha}=-\alpha\frac{\left((x_3-1)\vet k_3+\vet r_2\right)}{|\vet r-\vet k_3|^{\alpha+2}},
$$
$$
\vec{\nabla}_3 g_m=\vec{\nabla}_2g_m=m\frac{\hat\vet r_2}{r_2}g_m,
$$
$$
2\vec{\nabla}_3 g_m\cdot\vec{\nabla}_3\frac{1}{|\vet r-\vet k_3|^\alpha}=-2\alpha m\frac{g_m}{|\vet r-\vet k_3|^{\alpha+2}},
$$
$$
\nabla_3^2\frac{1}{|\vet r-\vet k_3|^\alpha}=-\frac{\alpha(\alpha-1)}{|\vet r-\vet k_3|^{\alpha+2}}.
$$
Thus we have
\be{lapfm}
\nabla_3^2 f_m=-\alpha(2m+1-\alpha)\frac{g_m}{|\vet r-\vet k_3|^{\alpha+2}}=0.
\ee
There is a non-trivial solution for $\alpha\ne0$ in \equ{lapfm}, that is $\alpha=2m+1$. Then 
\be{sol3d}
f_m=\frac{r_2^m e^{im\phi}}{|\vet r-\vet k_3|^{2m+1}}=\frac{r^m\sin^m\!\theta e^{im\phi}}{(1+r^2-2r\cos\theta)^{\frac{2m+1}{2}}},
\ee
is a solution of the Laplace equation in $d=3$ dimension. To obtain separable solutions, let us take the   $m$ derivatives with respect to $x=\cos\theta_d$ in \equ{ultra-esf}:
\be{mder}
\frac{\alpha(m,d)r^m}{(1+r^2-2rx)^{\frac{d+2m-2}{2}}}=\sum_{l=0}^\infty r^l \frac{d^m P_{l,d}(x)}{dx^m},\quad x=\cos\theta_d,
\ee
where $\alpha(m,d)=(d-2)d(d+2)\cdots (d+2m-4)$. Taking $d=3$ in \equ{mder} we obtain
\be{mder3}
\frac{\alpha(m,3)r^m}{(1+r^2-2rx)^{\frac{2m+1}{2}}}=\sum_{l=0}^\infty r^l \frac{d^m P_{l,3}(x)}{dx^m},\quad x=\cos\theta, r<1.
\ee
Substituting \equ{mder3} in \equ{sol3d} we conclude that
\be{sum3}
\sum_{l=0}^\infty r^l \sin^m\!\theta\frac{d^m P_{l}(x)}{dx^m} e^{im\phi}, x=\cos\theta, r<1, P_{l}=P_{l,3}
\ee
is a solution of the Laplace equation in $d=3$. Let us define the associated Legendre function by
\be{lf3}
P_l^m(\cos\theta)=\sin^m\!\theta \frac{d^m P_{l}(x)}{dx^m},\quad x=\cos\theta.
\ee
Note that since $P_l(\cos\theta)$ are polynomials of degree $l$, we have that $0\le m\le l$.

 By applying the laplacian operator in the equation \equ{sum3} and taking the account the linear independence of $r^l$ we have
\be{ale}
\hat{L}_3\left(P_l^m(\cos\theta)e^{im\phi}\right)=-l(l+1)P_l^m(\cos\theta)e^{im\phi}.
\ee
Writing $\hat{L}_3$ in the explicit form, we have the equation satisfied by the $P_l^m(\cos\theta)$:
\be{al3}
\frac{1}{\sin\theta}\frac{d}{d\theta}\left(\sin\theta\frac{dP_l^m}{d\theta}\right)+\left(l(l+1)-\frac{m^2}{\sin^2\!\theta}\right)P_l^m=0.
\ee
This is the associated Legendre equation. It is a Sturn-Liouville equation and this functions $P_l^m$ are orthogonal in $0\le\theta\le\pi$ with a weight $\sin\theta$ for a fixed $m$:
\be{ortplm}
\int_0^\pi \sin\theta P_l^m P_{l'}^m d\theta=0, \quad l\ne l'.
\ee 
Since that $e^{im\phi}$ are orthogonal in $[0,2\pi]$ we have a basis of functions that are orthogonal in $[0,2\pi]\times [0,\pi]$. We have $2l+1$ linearly indepependent functions for a given $l$, since $P_l^m(\cos\theta)e^{-im\phi}$ also satisfies \equ{ale}. 

In order to obtain the general eigenfunctions of $\hat{L}_d$, let us suppose by hypothesis that  a solution of the Laplace equation in $(d-1)$-dimension is given by
\be{sold-1}
g_{m_{d-2}\cdots m_1}=r_{d-1}^{m_{d-2}}\Psi_{m_{d-2}\cdots m_1}(\theta_{d-1}\cdots,\theta,\phi), \quad \theta_3=\theta,\theta_2=\phi,
\ee
where $\Psi_{m_{d-2}\cdots m_1}(\theta_{d-1}\cdots,\theta,\phi)$ is a function of the angles of $(d-1)$-dimension only.
For example for $d=3$ a solution in $3-1=2$ is the form $g_m=r_2^m e^{im\phi}$. Let us try a solution like \equ{fm}:
\be{triald}
f_{m_{d-2}\cdots m_1}=\frac{g_{m_{d-2}\cdots m_1}}{|\vet r-\hat{\vet k}_d|^\alpha}.
\ee
By the hypothesis
$$
\nabla^2_d g_{m_{d-2}\cdots m_1} =\nabla^2_{d-1}g_{m_{d-2}\cdots m_1} =0,
$$
$$
\vet r_{d-1}\cdot \vec{\nabla}_{d-1}g_{m_{d-2}\cdots m_1}=m_{d-2}g_{m_{d-2}\cdots m_1},
$$
$$
\nabla^2_d f_{m_{d-2}\cdots m_1}=0 \Rightarrow\frac{\alpha(\alpha-2m_{d-2}-d+2)}{|\vet r-\hat{\vet k}_d|^{\alpha+2}}g_{m_{d-2}\cdots m_1}=0
$$
For $\alpha\ne 0$ we have a non-trivial solution with $\alpha=2m_{d-2}+d-2$:
\be{sold}
f_{m_{d-2}\cdots m_1}=\frac{r_{d-1}^{m_{d-2}}\Psi_{m_{d-2}\cdots m_1}}{|\vet r-\hat{\vet k}_d|^{2m_{d-2}+d-2}}=\frac{r^{m_{d-2}}\sin^{m_{d-2}}\!\theta_d\Psi_{m_{d-2}\cdots m_1}}{|\vet r-\hat{\vet k}_d|^{2m_{d-2}+d-2}}
\ee
Now using \equ{mder} with $m=m_{d-2}$ we conclude that 
\be{fmsold}
\sum_{l=0}^\infty r^l \sin^{m_{d-2}}\!\theta_d\frac{d^{m_{d-2}} P_{l,d}(x)}{dx^{m_{d-2}}}\Psi_{m_{d-2}\cdots m_1}(\theta_{d-1}\cdots,\theta,\phi), \quad x=\cos\theta_d, r<1,
\ee
is a solution of the Laplace equation in $d$-dimension. 

We define like \equ{lf3} the ultraspherical associated Legendre function by
\be{lfd}
P_{l,d}^m(\cos\theta_d)=\sin^m\!\theta_d \frac{d^mP_{l,d}(x)}{dx^m}, \quad x=\cos\theta_d.
\ee
Applying recursively for $d-2$ until $d=3$, we have the form of the angular function $\Psi$:
\be{gensold}
\Psi_{l m_{d-2}\cdots m_2 m_1}= P_{l,d}^{m_{d-2}}(\cos\theta_d)P_{m_{d-2},d-1}^{m_{d-3}}(\cos\theta_{d-1})\cdots P_{m_2}^{m_1}(\cos\theta)e^{im_1\phi},
\ee
where $0\le m_{d-2}\le l, \quad 0\le m_{d-3}\le m_{d-2}, \cdots \quad 0\le m_1\le m_2$. 

Taking the laplacian operator in \equ{fmsold} and the linearly independence of $r^l$ we have
\be{angulard}
\hat{L}_d\Psi_{l m_{d-2}\cdots m_2 m_1}=-l(l+d-2)\Psi_{l m_{d-2}\cdots m_2 m_1}.
\ee
To obtain the equation satisfied by the $P_{l,d}^m(\cos\theta_d)$, we make $m_{d-2}=m$ and $m_1=m_2=\cdots m_{d-3}=0$ and use the explicit form of $\hat{L}_d$:
\be{equltrad}
\frac{1}{\sin^{d-2}\!\theta_d}\frac{d}{d\theta_d}\left(\sin^{d-2}\!\theta_d\frac{dP_{l,d}^m}{d\theta_d}\right)+\left(l(l+d-2)-\frac{m(m+d-3)}{\sin^2\!\theta_d}\right)P_{l,d}^m=0.
\ee
Note that for $d=3$ this equation reduces to equation of the associated Legendre functions. This equation is a Sturn-Liouville type and the $P_{l,d}^m$ are orthogonal with weight $\sin^{d-2}\!\theta_d$ in the interval $[0,\pi]$ for a fixed $m$:
\be{dortog}
\int_0^\pi\sin^{d-2}\!\theta_d P_{l,d}^m(\cos\theta_d)P_{l',d}^m(\cos\theta_d)d\theta_d=0, \quad l\ne l'.
\ee
Now using \equ{gensold} we have the orthogonality for the eigenfunctions $\Psi_{l m_{d-2}\cdots m_2 m_1}:$
\be{dortspi}
\int_\Omega\Psi_{l m_{d-2}\cdots m_2 m_1}\Psi^\ast_{l' m'_{d-2}\cdots m'_2 m'_1}d\Omega=0, \quad l\ne l', m_{d-2}\ne m'_{d-2}, \cdots m_1 \ne m'_1,
\ee 
where $d\Omega=\sin^{d-2}\!\theta_d\sin^{d-3}\!\theta_{d-1}\cdots\sin\theta d\theta_d d\theta_{d-1}\cdots d\theta d\phi$.

We have a set of mutually orthogonal functions of entire domain of the angles. The number of this functions for a given $l$ is
\be{numberoffunc}
N_d(l)=\frac{(d+2l-2)(d+l-3)!}{(d-2)!l!}.
\ee
\section{The Laplace equation solution with radial frontier condition}
Using the results of the previous sections we conclude that the general solution of the Laplace equation in $d$-dimension with radial frontier condition and finite for all value of the angles is
\be{solucaogeral}
\Phi(r,\Omega)=\sum_{l=0}^\infty\sum_{m_{d-2}=0}^l\cdots \sum_{m_{2}=0}^{m_3}\sum_{m_1=0}^{m_2}F_{l m_{d-2}\cdots  m_1}(r) \Psi_{l m_{d-2}\cdots m_1}(\Omega)+C.C,
\ee
where 
\be{radialF}
F_{l m_{d-2}\cdots  m_1}(r)=A_{l m_{d-2}\cdots  m_1}r^l+B_{l m_{d-2}\cdots m_1}r^{-(l+d-2)},
\ee
$A_{l m_{d-2}\cdots  m_1}$, $B_{l m_{d-2}\cdots  m_1}$ are complex constants to be determined by the frontier conditions and $C.C$ is the complex conjugated.

To write \equ{solucaogeral} in a compact form, let us define the ultraspherical harmonics:
\begin{eqnarray}
Y_{l m_{d-2}\cdots m_1}=f_{l m_{d-2}\cdots m_1}\Psi_{l m_{d-2}\cdots m_1},\\
Y_{l m_{d-2}\cdots -m_1}=f_{l m_{d-2}\cdots m_1}\Psi_{l m_{d-2}\cdots m_1}^{\ast},\label{norm}
\end{eqnarray}
where $f_{l m_{d-2}\cdots m_1}$ is a normalization coefficient such that
\be{ylmort}
\int_\Omega Y_{l m_{d-2}\cdots m_1}Y_{l' m_{d-2}'\cdots m_1'}^\ast d\Omega=\delta_{l l'}\delta_{m_{d-2} m_{d-2}'}\cdots \delta_{m_2 m_2'} \delta_{m_1 m_1'},
\ee
and are given by
\be{factor}
f_{l m_{d-2}\cdots m_1}= N^{(d)}_{l m_{d-2}} N^{(d-1)}_{m_{d-2} m_{d-3}}\cdots N^{(3)}_{m_2 m_1},
\ee
where $N^{(d)}_{n m}$ is the normalization factor of the ultraspherical associated Legendre function in $d$-dimension:
\be{nfactor}
\left(N^{(d)}_{n m}\right)^2\int_0^\pi\sin^{d-2}\!\theta_d \left(P_{n,d}^m(\cos\theta_d)\right)^2 d\theta_d=1,
\ee
\be{normn}
N^{(d)}_{nm}=\sqrt{\frac{(2n+d-2)}{(d-2)}\frac{\Omega_{d-1}}{\Omega_d}\frac{(d-3)!(n-m)!}{(d+n+m-3)!}}.
\ee
Now \equ{solucaogeral} can be written as
\be{solg}
\Phi(r,\Omega)=\sum_{l=0}^\infty\sum_{m_{d-2}=0}^l\cdots \sum_{m_{2}=0}^{m_3}\sum_{m_1=-m_2}^{m_2} F_{l m_{d-2}\cdots  m_1}(r) Y_{l m_{d-2}\cdots m_1}(\Omega).
\ee

\section*{Acknowledgements}
The {\it Conselho Nacional de Desenvolvimento Cientifico e Tecnologico},
$CNPq$-Brazil is gratefully acknowledged for the financial support.
\vskip1cm
{\bf\noindent This paper is dedicated to the love of my life: my wife Isabel Mara.}
\appendix
\section{The volume element in ultraspherical coordinates}
We see from the definition of the ultraspherical coordinates that 
\begin{eqnarray}
x_j=r_j\cos\theta_j, \quad r_{j-1}=r_j\sin\theta_j,\quad 0\le \theta_j \le\pi, \quad j\ge 3,\label{a-ultrad}\\
x=\rho\cos\phi,\quad y=\rho\sin\phi, \quad 0\le\phi\le 2\pi, \quad \rho=r_2.
\end{eqnarray}
The volume element in $d=2$ is $dV_2=dxdy=\rho d\rho d\phi$. In $d=3$ we can write 
\be{vol3}
dV_3=dx_3dV_2=\rho dx_3d\rho  d\phi.
\ee
Now using \equ{a-ultrad} with $j=3$,  $x_3=r_3\cos\theta_3$, $\rho=r_2=r_3\sin\theta_3$ we have that $dx_3d\rho=r_3dr_3d\theta_3$. Then
$$
dV_3=(r_3\sin\theta_3)r_3dr_3d\theta_3 d\phi=r_3^2\sin\theta_3 dr_3 d\theta_3 d\phi=r_3^2 dr_3 d\Omega_3
$$
Now supposing that the volume element in $(d-1)$-dimension is 

\be{a-vold1}dV_{d-1}=r^{d-2}_{d-1}dr_{d-1}d\Omega_{d-1},\ee we have
\be{a-vold}
dV_d=dx_d dV_{d-1}=r^{d-2}_{d-1}dx_d dr_{d-1}d\Omega_{d-1}.
\ee
Now using \equ{a-ultrad} with $j=d$, $x_d=r_d\cos\theta_d$, $r_{d-1}=r_d\cos\theta_d$, we get $dx_d dr_{d-1}=r_d dr_d d\theta_d$. Hence 
\be{a-vold2}
dV_d=r_d^{d-1}\sin^{d-2}\!\theta_d dr_d d\theta_d d \Omega_{d-1}=r_d^{d-1} dr_d d\Omega_d,
\ee
where
\be{a-an}
d\Omega_d=\sin^{d-2}\!\theta_d d\theta_d d\Omega_{d-1}.
\ee
Using \equ{a-an} recursively, we conclude that
\be{a-and}
d\Omega_d=\sin^{d-2}\!\theta_d\sin^{d-3}\!\theta_{d-1}\cdots \sin\!\theta_3 d\theta_d d\theta_{d-1}\cdots d\theta_3 d \phi.
\ee
The solid angle $\Omega_d$ is given by
\be{solidan}
\Omega_d=\left(\int_0^\pi \sin^{d-2}\!\theta_d d\theta_d\right) \Omega_{d-1}=\sqrt{\pi}\frac{\Gamma(\frac{d-1}{2})}{\Gamma(\frac{d}{2})}\Omega_{d-1}=2\frac{\pi^{\frac{d}{2}}}{\Gamma(\frac{d}{2})}.
\ee
\section{The normalization factor of $P_{l,d}^n$}

From the definition of the ultraspherical polynomials we have 
\be{b-gener}
\frac{1}{|\vet r-\vet r'|^{d-2}}=\sum_{l=0}^\infty \frac {r_{<}^l}{r_{>}^{l+d-2}}P_{l,d}(\cos\gamma_d),
\ee
where 
$\cos\gamma_d=\hat{\vet r}\cdot\hat{\vet r}'=\cos\theta_d\cos\theta_d'+\sin\theta_d\sin\theta_d'\cos\gamma_{d-1}$ and $r_< (r_>)$ is the smaller (larger) of $r$ and $r'$.
Taking the laplacian in both sides of \equ{b-gener} we obtain
\be{b-diracd}
(2-d)\Omega_d\delta^d(\vet r-\vet r')=\sum_{l=0}^\infty \left(\hat{O}_d(r)\frac {r_{<}^l}{r_{>}^{l+d-2}}-\frac{l(l+d-2)}{r^2}\frac {r_{<}^l}{r_{>}^{l+d-2}}\right)P_{l,d}(\cos\gamma_d),
\ee
where $\delta^d(\vet r-\vet r')$ is the Dirac delta distribution in $d$-dimension,
$$\delta^d(\vet r-\vet r')=\frac{\delta(r-r')}{r^{d-1}}\frac{\delta(\theta_d-\theta_d')}{\sin^{d-2}\!\theta_d}\cdots\frac{\delta(\theta_3-\theta_3')}{\sin\theta_3}\delta(\phi-\phi'),$$
and $\hat{O}_d(r)$ is the radial operator,
$$
\hat{O}_d(r)=\frac{1}{r^{d-1}}\dpa{}{r}\left(r^{d-1}\dpa{}{r}\right).
$$
Multiplying \equ{b-diracd} by $r^{d-1}$ and integrating from  $r'-\epsilon$ to $r'-\epsilon$ with $\epsilon\rightarrow0$, we have
\be{b-complet1}
(d-2)\Omega_d\frac{\delta(\theta_d-\theta_d')}{\sin^{d-2}\!\theta_d}\cdots\frac{\delta(\theta_3-\theta_3')}{\sin\theta_3}\delta(\phi-\phi')=\sum_{l=0}^\infty (2l+d-2)P_{l,d}(\cos\gamma_d).
\ee
Since the ultraspherical harmonics are orthogonal in the domain of all angles we can write
\be{b-pexp}
P_{l,d}(\cos\gamma_d)=\sum_{m_{d-2}=0}^l\cdots \sum_{m_1=-m_2}^{m_2} A_{l m_{d-2}\cdots  m_1} Y_{l m_{d-2}\cdots m_1}(\Omega_d).
\ee
Using the orthogonality condition \equ{ylmort} and \equ{b-complet1} with \equ{b-pexp} we obtain
$$
A_{l m_{d-2}\cdots  m_1}=\frac{(d-2)\Omega_d}{2l+d-2}Y^\ast_{l m_{d-2}\cdots  m_1}(\Omega_d').
$$
Thus we obtain the ultraspherical harmonics addition:
\be{b-add-harm}
P_{l,d}(\cos\gamma_d)=\frac{(d-2)\Omega_d}{2l+d-2}\sum_{m_{d-2}=0}^l\cdots \sum_{m_1=-m_2}^{m_2}Y_{l m_{d-2}\cdots m_1}(\Omega_d)Y^\ast_{l m_{d-2}\cdots  m_1}(\Omega_d').
\ee
We can write the equation above in a useful formula
\be{b-add2}
P_{l,d}(\cos\gamma_d)=K_{l,d}\sum_{m_{d-2}=0}^l (2m_{d-2}+d-3)(N^{(d)}_{lm_{d-2}})^2P_{l,d}^{m_{d-2}}(\cos\theta_d)P_{l,d}^{m_{d-2}}(\cos\theta_d')P_{m_{d-2},d-1}(\cos\gamma_{d-1}),
\ee
where $$K_{l,d}=\frac{\Omega_d}{\Omega_{d-1}}\frac{(d-2)}{(2l+d-2)(d-3)}.$$

Now deriving the equation \equ{b-add2} with respect to $\theta_d'$ $n$-times with  $\theta_d'=0$ and equalling the independent terms we have
\be{NLM}
N^{(d)}_{ln}=\frac{1}{\sqrt{(2n+d-3)K_{l,d}P^{'(n)}_{l,d}(1)P^{'(n)}_{n,d-1}(1)}},
\ee
where $P^{'(n)}_{l,d}(1)$ and $P^{'(n)}_{n,d-1}(1)$ is the $n$-th derivative of $P_{l,d}(x)$ and $P_{n,d-1}(x)$ with respect to $x$ evaluated in $x=1$. This can be obtained from \equ{mder} with $x=1$:
\be{p1}
P^{'(n)}_{l,d}=\frac{(d-2)d(d+2)\cdots (d+2n-4)(d+n+l-3)!}{(l-n)!(d+2n-3)!}.
\ee
Then
$$
P^{'(n)}_{l,d}(1)P^{'(n)}_{n,d-1}(1)=\frac{(d+n+l-3)!}{(d-4)!(l-n)!(d+2n-3)}.
$$
We finally obtain
\be{NF}
N^{(d)}_{ln}=\sqrt{\frac{(2l+d-2)}{(d-2)}\frac{\Omega_{d-1}}{\Omega_d}\frac{(d-3)!(l-n)!}{(d+l+n-3)!}}
\ee
\end{document}